# Surface plasmon polaritons on rough metal surfaces: Understanding the formation of laser induced periodic surface structures in metals

Y. Fuentes-Edfuf,[1] J. A. Sánchez-Gil,[2,*] C. Florian,[1] V. Giannini,[2] J. Solis,[1] and J. Siegel[1,†]

1. Laser Processing Group, Instituto de Optica (IO-CSIC), Consejo Superior de Investigaciones Científicas,CSIC, Serrano 121, 28006 Madrid, Spain

2. Instituto de Estructura de la Materia (IEM-CSIC), Consejo Superior de Investigaciones Científicas, Serrano 121, 28006 Madrid, Spain

* *j.sanchez@csic.es*, † *j.siegel@csic.es*

**ABSTRACT:** The formation of self-organized laser induced periodic surface structures (LIPSS) in metals, semiconductors and dielectrics upon pulsed laser irradiation is a well-known phenomenon, receiving increased attention due to its huge technological potential. For the case of metals, a major role in this process is played by surface plasmon polaritons (SPPs) propagating at the interface of the metal with the medium of incidence. Yet, simple and advanced models based on SPP propagation sometimes fail to explain experimental results, even of basic features as the LIPSS period. We experimentally demonstrate, for the particular case of LIPSS on Cu, that significant deviations of the structure period from the predictions of the standard model are observed, which are very pronounced for elevated angles of laser incidence. In order to explain this deviation, we introduce a model based on the propagation of a SPP on a rough surface that takes into account the influence of the specific roughness properties on the SPP wave vector. Good agreement of the modelling results with the experimental data is observed, which highlights the potential of this model for the general understanding of LIPSS in other metals.

Table of Contents artwork

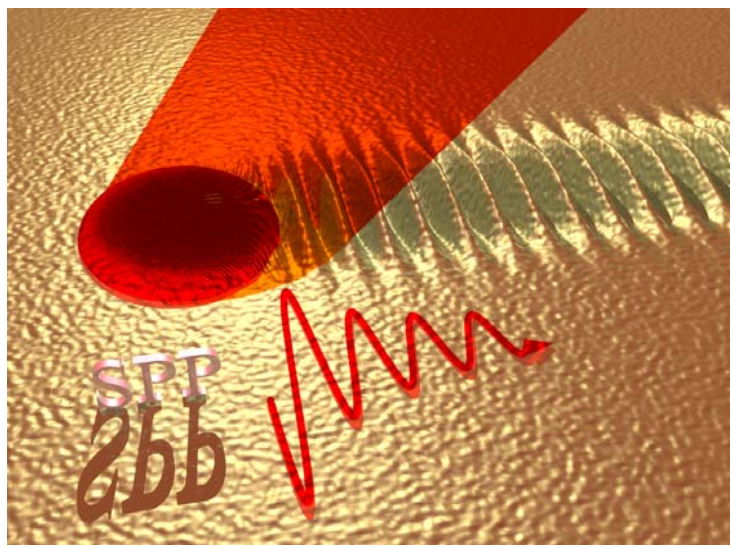

INTRODUCTION:

Laser Induced Periodic Surface Structures (LIPSS) are formed when a material is exposed to multiple laser pulses at energies above the modification threshold.[1–4] The predominant mechanism giving rise to LIPSS formation is related to the interference of incident laser light with a wave propagating at the surface, leading to a periodic intensity modulation that is imprinted into the material.[5] LIPSS manifest in metals,[6–8] semiconductors[9–11] and dielectrics,[12–16], both organic and inorganic, and can feature different sizes, shapes and orientations that are determined by a complex interplay of laser irradiation and material parameters. The enormous richness of structures that can be fabricated in that way in viturally all materials with countless applications, combined with the inherent simplicity offered by a single-beam laser fabriaction approach based on self-organization, have triggered intense research over the past decades[3,17–21] Amongst the irradiation parameters, the angle of incidence $\theta$ of the laser beam occupies a prominent role, since it directly influences the period of the so-called Low Spatial Frequency (LSF) LIPSS, consisting of parallel ripples aligned perpendicular to the laser polarization. Despite the importance of $\theta$, there are surprisingly few works on systematic studies of this parameter, and these works are limited to a very narrow selection of materials.[2,22–25]

For the specific case of metals, the wave propagating at the surface that contributes to ripple formation is generally identified as a surface plasmon polariton[26–29] coupled in from the incident light with help from a non-negligible surface roughness. The role of surface roughness is thus to impart momentum to the incident wave, bridging the momentum mismatch between the parallel component of the wavevector of such incident wave with that of the SPP, which lies beyond the light cone.[30,31] In fact, the surface roughness parameters play a crucial role in determining the light-SPP coupling efficiency.[32,33] The emergence of LIPSS is associated to the formation of a surface wave as a result of the interference between the incident field and the excited SPP, yielding the well-known expression of the LIPSS period.[34] In general, though, the role of surface roughness is assumed to be only to allow SPP excitation, and the calculated SPP wavevector used corresponds to that at a planar dielectric-metal interface. This approach does reproduce the general tendency of the LIPPS period as a function of angle of incidence, but fails to predict correctly the period values.[22]

In this regard, it should be recalled that SPP propagating along a rough surface can also couple into outgoing waves.[35] More importantly for LIPSS formation, and often overlooked, is the fact that the SPP itself senses the surface roughness along its propagation. Attempts to account for this discrepancy have been carried out by considering the rough metal surface in form of an additional thin layer composed of air and metal, whose dielectric function is modelled by Maxwell-Garnett theory of effective media, slightly improving the agreement with experimental results.[22,29] Nonetheless, the surface roughness appears in this model as the metal filling fraction of such intermediate air-metal interface, without including in detail surface roughness statistical parameters.

Here, we report a strong deviation of the angle dependence of the LIPSS period on Cu from the predictions of all existing models based on SPP propagation. In order to explain the observed differences, we introduce a model based on the modification of the propagation of surface plasmon polaritons on a rough surface that incorporates the strong dependence of the SPP wavevector on surface roughness parameters.[36,37] Upon introducing such specific roughness parameters determined experimentally into the rough-SPP model, we show that the experimental LIPSS periods can be properly obtained by the model.

RESULTS:

Laser irradiation experiments have been performed by exposing the Cu sample to a train of focused ultrashort laser pulses while moving the sample at different speeds. Given the numerous experimental parameters that influence the formation of LIPSS in metals even for constant laser pulse properties (800 $nm$ wavelength, 100 $fs$ pulse duration, p-polarization, 100 $Hz$ repetition rate), we have performed several independent irradiation experiments in order to narrow down the global parameter space. First, we have explored the parameter space given by the laser fluence $F$ and the effective pulse number $N_{eff}$ in order to identify the optimum conditions for the formation of LIPSS for a constant angle of incidence ($\theta = 0°$). We found that the minimum pulse number necessary for the appearance of reasonably well-pronounced and aligned LIPSS in Cu was $N_{eff} = 1000$ (cf. experimental section for the definition of $N_{eff}$), which is consistent with other works, reporting similarly high values.[38] The fluence range for LIPSS to appear under these conditions was $F_{eff} = [15, 35]\ mJ/cm^2$ (cf. experimental section for the definition of $F_{eff}$).

In a second step, we varied the angle of incidence while keeping $N_{eff}$ and $F_{eff}$ constant. Figure 1(a-f) shows SEM images of the resulting LIPSS for three angles of incidence ($\theta = 0°, 30°$ and $52°$). Several observations can be made: First, the periodic structures are aligned perpendicularly to the laser polarization and their period $\Lambda$ increases with the angle of incidence, as expected for low spatial frequency (LSF) LIPSS (so-called ripples) in metals. Second, for a given angle, two different ripple periods are observed in different regions of the laser-written track. This is particularly evident for the results shown for $\theta = 52°$ (cf. Figure 1c,f), featuring a shorter period towards the center of the track (inner LIPSS) and a longer one at the border (outer LIPSS). It should be reminded that under the present laser excitation conditions with a Gaussian-shaped intensity profile, the center of the track has experienced a higher local fluence than the border. This position-dependent, and thus fluence-dependent period is astonishing and will be one of the main foci of interest in the present paper.

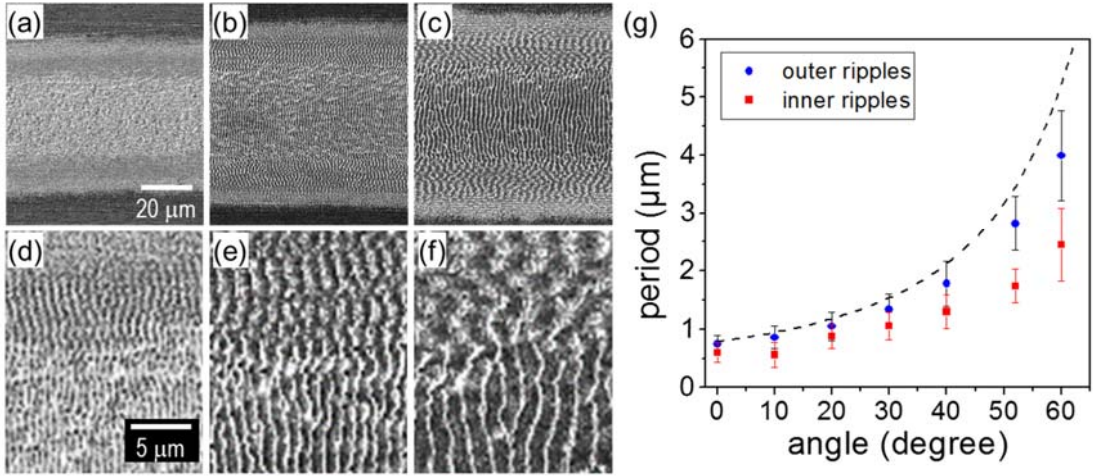

*Figure 1: Angle dependence of the ripple period of LSF-LIPSS in Cu. (a-c) SEM images of laser-written tracks at $F_{eff} = 20\ mJ/cm^2$ and $N_{eff} = 1000$ for (a) $\theta = 0°$, (b) $\theta = 30°$ and (c) $\theta = 52°$. (d-f) Higher magnification images of the corresponding regions marked in (a-c) of the transition region of the inner ripples (track center) to outer ripples (track border). (g) Dependence of the ripple period for the two types of ripples with the angle of incidence. The blue*

*circles and red squares correspond to experimental data, whereas the dashed curve represents the result of the simple plasmonic model (see text).*

Indeed, it is well known that a change of fluence can lead to different classes of LIPSS in metals, ranging from high spatial frequency (HSF)-LIPSS and LSF-LIPSS at low and moderate fluence, to so-called grooves and spikes at high fluence.[39] These LIPSS classes differ strongly from each other in terms of their spatial features, with HSF-LIPSS having a period of less than half the laser wavelength and being aligned parallel to the laser polarization, LSF-LIPSS having a period similar to the laser wavelength at normal incidence and being aligned perpendicular to the polarization. Grooves and spikes have both a size much larger than the laser wavelength. Comparing these known LIPSS classes to our case of two types of ripples sharing the same orientation (Figure 1a-f) leads us to the conclusion that both belong to the same LIPSS class, namely LSF-LIPSS. We attribute the absence of other experimental works (to the best of our knowledge) reporting two types of LSF-LIPSS to the fact that only few studies at all have explored the angle dependence of LIPSS. For zero angle the difference in period is small and close to the experimental resolution, as opposed to elevated angles, where the two types can clearly be distinguished.

In order to investigate the origin of the two types of LIPSS in more detail we have extended the study of the angle dependence to cover a range from 0° to 60°. The results are plotted in Figure 1(g). The most striking observation is that two different periods are present for a single angle, confirming the observations made in Figure 1(a-b). While it can be argued that for small angles the difference in period is negligible, at larger angles the difference is considerable. We have included in Figure 1(c) data for a period $\Lambda^{s+}$ calculated from simple plasmonic theory,[34,40] based on the propagation of a surface plasmon polariton at the air-metal interface, according to

$$\Lambda^{s+} = \frac{\lambda}{\text{Re}(\tilde{n}) - \sin(\theta)} \quad (1)$$

with $\lambda$ being the laser wavelength. $\text{Re}(\tilde{n})$ is typically assumed to be the real part of the complex refractive index of the interface, essentially proportional to the SPP wavevector $k_{SPP}^0$ on a planar air-metal interface, so that

$$\tilde{n} = k_{SPP}^0 \cdot \lambda/(2\pi) = [(\epsilon_{air} \cdot \epsilon_{metal})/(\epsilon_{air} + \epsilon_{metal})]^{1/2} \quad (2)$$

with $\epsilon_{air}$ and $\epsilon_{metal}$ being the dielectric functions of air and the metal. It can be seen that the general trend of both periods as a function of $\theta$ does indeed follow the theory, but the absolute values differ considerably for both types of LIPSS observed experimentally, being systematically lower in both cases. It should be noted here that an additional subwavelength period $\Lambda^{s-}$ is predicted by a variation of Eq. (1), in which the minus sign is replaced by a plus sign. However, in the present case of Cu, we have not found experimental evidence for this period, whereas we did observe it in our recent work on silicon.[24,25]

A deviation of experimentally observed period values from values predicted by the simple plasmonic model as observed in Figure 1(g) is often reported in literature and several modifications to the model have been proposed. The most widely accepted one is based on a transient change of the dielectric function of the material *during* laser irradiation, leading to a transient increase in the free electron density.[27] While this correction is vital for explaining the behavior in semiconductors and dielectrics, for the case of metals with already a high density of free electrons the expected change is only minor. Another addition to the plasmonic model is based on taking into account the hydrodynamics taking place after laser-induced melting.[9] Whilst this modification does explain the additional formation of the so-called groove structures, with larger size and parallel orientation to the laser polarization, it does not yield a significant change of the period of the LSF-LIPSS with respect to the other theories. A further modification to the simple plasmonic model is based on the fact that LSF-LIPSS often feature a nanostructure superimposed to them, in form of randomly distributed nanoparticles. Hwang et al.[22] argue that this fine structure leads to an effective change of the dielectric constant $\epsilon_{air}$ and directly influences SPP propagation. The authors introduce this concept in their model in form of an air-nanostructure composite layer of a certain thickness and fill factor, which is then described by the Maxwell–Garnett theory of effective media. They implement this concept by assigning an effective dielectric function $\epsilon_{eff}$ to the air-nanostructure composite, replacing $\epsilon_{air}$ used in the expression for calculating ñ. Using this approach, the authors were able to fit their experimental data by adjusting the volume fraction of the metal inclusions in the composite "air layer". In a recent work the same group were able to determine the volume fraction of the metal inclusions by means of AFM measurements.[29]

An important point that had been neglected in Refs.[22,29] is the fact that the nanostructure features used for the model input were those of the final structures. However, the relevant nanostructure for surface plasmon propagation to occur is the one present at the threshold of LIPSS formation. The formation of LIPSS is based on multiple pulse irradiation during which the surface topography and morphology evolves as the pulse number increases. During the first few pulses the surface roughness increases homogeneously, until from a certain pulse number onwards LIPSS begin to form. Thus, it is the surface roughness at this moment, which is relevant for calculation of SPP coupling and propagation.

A simple way to investigate the progressive formation of the two types of ripples observed as the pulse number increases is to write single tracks at different values of $N_{eff}$. Figure 2 shows a set of SEM images of track regions written at constant fluence ($F_{eff} = 25\ mJ/cm^2$) and angle ($\theta = 52°$), with $N_{eff} = 5, 10, 50, 100, 1000$. While at $N_{eff}$ = 5, 10 the track manifests as a slight brightening of the surface, for $N_{eff} = 50$ and higher a significant corrugation of the surface can be appreciated. At $N_{eff} = 100$, the characteristic periodic structures of both, the inner and outer ripples already begin to emerge. From these results we conclude that an appropriate pulse number threshold for ripple formation is $N_{eff,thres} = 50$.

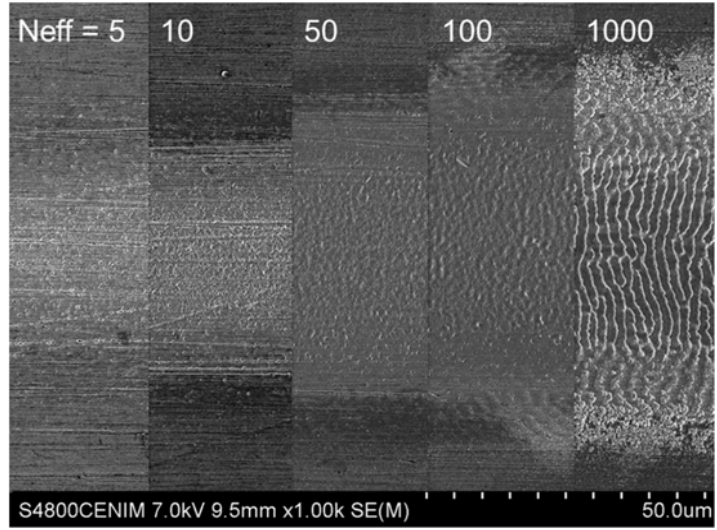

*Figure 2: Sequence of SEM images showing the evolution of the surface morphology of Cu as a function of pulse number $N_{eff}$ for $\theta = 52°$.*

Since SEM data provide only access to morphological information, AFM measurements need to be performed in order to obtain a quantitative study of the surface roughness, which is required for the model here proposed. Representative AFM maps recorded with a step size $d = 30\ nm$ in the center of the written track for different values of $N_{eff}$ are displayed in Figure 3(a-d). The unexposed surface features already a considerable roughness, with typical feature sizes of $100\ nm$, which remains essentially unchanged for $N_{eff} = 5$. A significant coarsening is observed at $N_{eff} = 50$, although no signs of vertically aligned ripples can be seen. Such ripples are clearly observed at $N_{eff} = 1000$, accompanied by a strong increase in modulation depth.

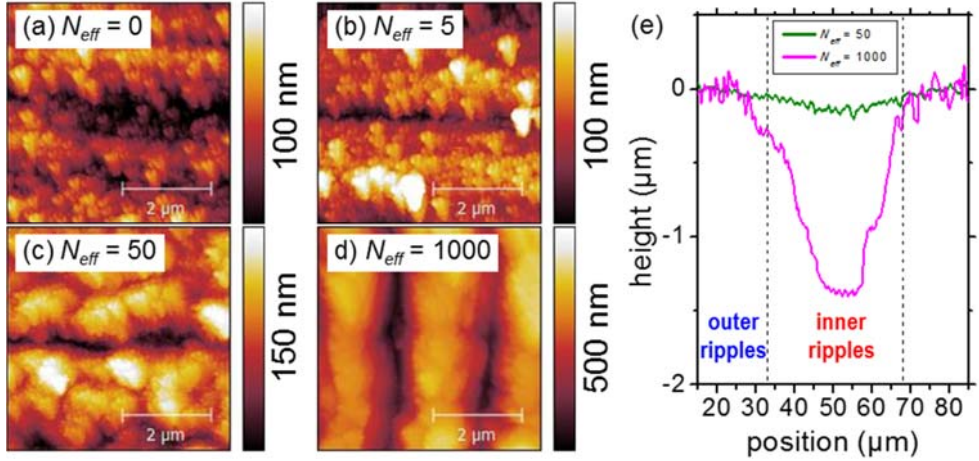

*Figure 3: (a-d) Sequence of AFM images recorded with a step size $d = 30\ nm$, showing the evolution of the surface topography of Cu as a function of pulse number $N_{eff}$ for $\theta = 52°$ at the center of the written track (cf. Figure 2). (e) Cross section of a laser written track with $N_{eff} = 50$ and $1000$ recorded with an AFM and step size $d = 600\ nm$. The dashed vertical lines mark the border between the different regimes of inner and outer ripples, extracted from the SEM images shown in Figure 2.*

Table 1 displays the extracted characteristic statistical data of the AFM measurements, namely the RMS height $\delta_{nano}$ of the roughness and the corresponding correlation length $\sigma_{nano}$. It can be seen that the nanoscale roughness of the unexposed sample is already relatively high and increases only slightly during the first 50 pulses, slightly more in the inner region.

| $N_{eff}$ | inner region | | outer region | |
|---|---|---|---|---|
| | $\delta_{nano}$ (nm) | $\sigma_{nano}$ (nm) | $\delta_{nano}$ (nm) | $\sigma_{nano}$ (nm) |
| 0 | 17 | 690 | 17 | 690 |
| 5 | 20 | 380 | 17 | 690 |
| 50 | 28 | 480 | 19 | 560 |
| 1000 | 121 | - | 115 | - |

*Table 1: Statistical data of AFM measurements of laser written tracks for different pulse numbers Neff at θ = 52°. $\delta_{nano}$ represents the RMS height and $\sigma_{nano}$ the correlation length.*

Figure 3(e) shows AFM cross sections of tracks written at different pulse numbers. The region in which the inner ripples are formed (between the dashed vertical lines in Figure 3(e), as determined from the SEM images shown in Figure 2) corresponds to the region of strong ablation with a maximum crater depth of ~1.3 μm at $N_{eff} = 1000$. In contrast, the outer ripples are formed in a region of weak ablation. It is worth noting that the AFM measurement of the track cross section in Figure 3(e) was performed over a 100 μm scale at a much larger step size ($d = 600\ nm$) than the one shown in Figure 3(a-d) (5 μm scale with $d = 30\ nm$). A comparison of both types of AFM measurements, of a small area with high resolution versus a large area with low resolution, reveals that the laser-induced surface roughness combines coarse structures superimposed with a fine structure, both for $N_{eff} = 50$ and $N_{eff} = 1000$. While the strongly hierarchical/multiscale morphology of LIPSS is well-known and opens interesting possibilities of fabricating novel surface structures with new funcionalities,[17,41] it poses an enormous challenge for our attempt to understand SPP propagation on such multiscale surfaces.

MODELLING:

We modeled our system based on the fact that the emergence of LIPSS is associated to the electromagnetic field modulations resulting from the interference between the incident field and the excited SPP. Then, the LIPSS period $\Lambda^{s+}$ - dependence on the incident angle $\theta$ can be obtained from this well-known expression[34,40] shown by Eq. (1) above. In our model, we will assume that the SPP itself is perturbed by the presence of surface roughness; as a result, its wavevector $k_{SPP}$ can be modified, on average, with respect to that for a planar surface $k_{SPP}^{(0)}$. Bearing in mind that surface roughness is not deterministic, accounting theoretically for the average dispersion relation of SPP on a randomly rough surface is a complex task. We will take advantage of the fact that surface roughness precursor of LIPSS is normally much smaller than the incoming wavelength, so that perturbation theories may hold.[36,37] In particular, we will make use of the expression Eq. (A42) derived in Ref.[37] for the roughness-induced modification of the SPP wavevector.

$$\Delta k_{SPP} = \frac{\delta^2 \sigma^2}{2} \frac{|\varepsilon_r|^{1/2}}{(\varepsilon_r + 1)^2} exp\left[-\left(k_{SPP}^{(0)}\right)^2/4\right]$$
$$\times \int_0^\infty kdk \frac{\alpha - \varepsilon\alpha_0}{k^2 - \left(k_{SPP}^{(0)}\right)^2} exp\left[-\frac{k^2\sigma^2}{4}\right] F\left(k, k_{SPP}^{(0)}, \sigma\right) \quad (3)$$

With $\varepsilon_{metal} = \varepsilon_r + i\varepsilon_i, |\varepsilon_r| \gg |\varepsilon_i|$, and $\alpha_0 = \left[k^2 - \left(\frac{\omega}{c}\right)^2\right]^{1/2}, \alpha = \left[k^2 - \varepsilon\left(\frac{\omega}{c}\right)^2\right]^{1/2}$. Therefore, the complex refractive index of the interface that has to be included in Eq. (1) to yield the LIPSS periods within our model, accounting for the roughness-induced modification of the SPP wavevector on a planar air-metal interface, is

$$ñ = \left(k_{SPP}^{(0)} + \Delta k_{SPP}\right) \cdot \lambda/(2\pi) \quad (4)$$

Therein, one has to consider surface roughness as a Gaussian-correlated, Gaussian statistics random process, characterized by the RMS height $\delta$ (assumed small compared to the incoming wavelength $\lambda, \delta \ll \lambda$, and by its correlation length $\sigma$. Recall that inverse methods based on light-scattering data can be in turn used to determine the surface autocorrelation functions.[42] The resulting modification of $\Delta k_{SPP}$ can be strongly dependent on surface roughness parameters.[37] The RMS height $\delta$ is accounted for up to second order, $\delta^2$. The impact of the correlation length $\sigma$ is more subtle, since it accounts for the lateral dimensions of roughness, which are in turn inversely proportional to the imparted transverse momentum; thus correlation lengths smaller than the SPP wavelength are usually required to affect more strongly the SPP wavevector. A sketch of the excitation scenario is shown in Figure 4.

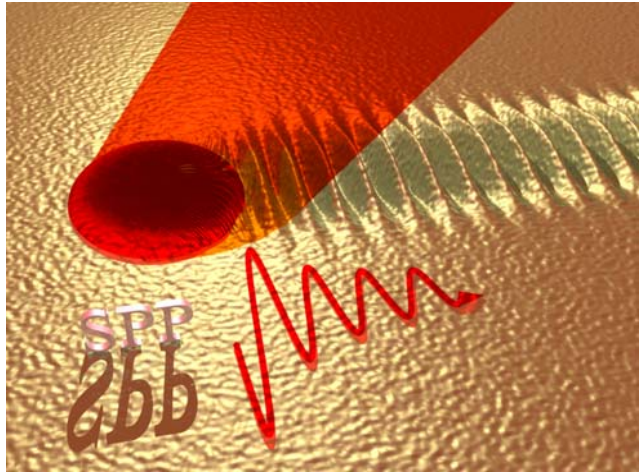

*Figure 4: Sketch of the excitation scenario of SPP, leading to the formation of LIPSS.*

As discussed before (cf. Figure 2), we now consider $N_{eff} = 50$ as the threshold for LIPSS formation ($N_{eff,thres}$). Without the presence of the additional coarse microstructure, the roughness parameters of the nanostructure listed in Table 1 should be appropriate as the relevant input parameter for our model. Yet in view of the existence of additional microscale structures, in particular the ablation crater in the region of the inner ripples, the values listed in the table are most likely underestimated. Because of this we have preferred to calculate the ripple periods by leaving the surface roughness parameter $\delta$ as fit parameter rather than to use the $\delta_{nano}$ values listed in Table 1. However, we do use the $\sigma_{nano}$ listed in Table 1, since the lateral dimensions at the nanoscale are well accounted for through AFM. As mentioned above, finer lateral dimensions result in higher momentum imparted into SPPs, thus having more impact than coarser lateral dimensions on the roughness-induced modification of the SPP wavevector [see Eq. (3)]. Figure 5 displays a plot of the period of the two types of ripples, combining the experimental data already shown in Figure 1 with the results of our model. It can be seen that the model fits the experimental data very well for all angles, much better than the simple plasmonic model not taking into account the surface roughness.

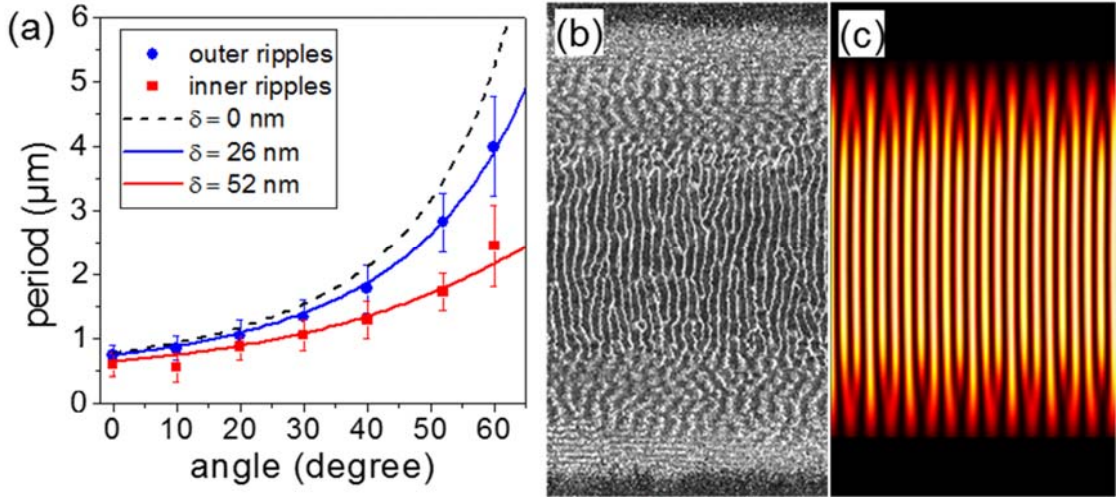

*Figure 5: (a) Dependence of the ripple period for the two types of ripples in Cu with the angle of incidence. The blue circles and red squares correspond to experimental data, whereas the solid curves represent the result of the plasmonic model taking into account the surface roughness. The dashed curve represents the result of the model assuming a perfectly flat surface. (b,c) Comparison of experimental and modelling results at $\theta = 52°$: (b) SEM image for $N_{eff} = 1000$. (c) Snapshot of field intensity distribution of SPPs propagating on areas of different surface roughness.*

The fitted roughness parameters used in these calculations are $\delta = 26\ nm$ ($\sigma_{nano} = 560\ nm$) for the outer ripples and $\delta = 52\ nm$ ($\sigma_{nano} = 480\ nm$) for the inner ripples. A comparison with the nanoscale roughness parameters ($\delta_{nano} = 19\ nm$, $\sigma_{nano} = 560\ nm$ for the outer ripples and $\delta_{nano} = 28\ nm$, $\sigma_{nano} = 480\ nm$ for the inner ripples) shows that a relatively good agreement is obtained for the outer ripples whereas a clear underestimation is obtained for the inner ripples. As explained before, we attribute this poor match to the contribution of microscale roughness, which is particularly important for the inner ripples due to the presence of an ablation trench.

In order to visualize in two dimensions the importance of the local surface roughness on the wavevector of the SPP we show in Figure 5(b,c) a direct comparison of experimental and modelling results for a fixed angle of incidence, $\theta = 52°$. For the model we have calculated the instantaneous field intensity distribution of SPPs propagating on a surface with a heterogeneous roughness, the central stripe having a higher roughness $\delta$ and the outer regions a lower roughness. This is done in practice by assuming that the SPP wavevector, given by Eqs. (3) and (4), is different in the inner and outer regions, as obtained with the different surface roughness parameters mentioned above; such SPP wavevectors are thus identical to those used to fit the LIPSS periods in Figure 5(a). Apart from the inherent local randomness observed in the real ripples, the agreement in Figure 5(b,c) is remarkable, giving further support to the assumptions made in our theoretical model.

**CONCLUSIONS:**

From an experimental point of view, we have observed the co-existence of two types of periodic surface structures in copper upon scanning the sample while exposing it to a pulse train of ultrashort laser pulses at an elevated angle of incidence. The structures can be distinguished by their relative position and period, with the longer period being located at the track border and the shorter period being in the track center. Our model based on the propagation of surface plasmon polarizations on rough surfaces demonstrates that the surface roughness strongly influences the period of the structures formed. The periods predicted by the model are consistent with the experimental results for roughness values observed experimentally, although the hierarchical/multiscale morphology of LIPSS would need to be taken into account in order to achieve a perfect match.

METHODS:

Irradiation experiments were performed using an amplified laser system for irradiation, providing pulses of $120\,fs$ FWHM (Full Width at Half Maximum) at $800\,nm$ central wavelength with a repetition rate of $100\,Hz$. The pulse energy is adjusted by a combination of a half-waveplate and a polarizing beam splitter cube. A second half-waveplate is used to set the polarization of the pulse to p-polarized at the sample plane. The beam passes through a beam shaping circular aperture with a diameter $\emptyset = 3.5\,mm$ before being focused by a lens with focal length $f = 150\,mm$ at the sample surface, at an angle of incidence $\theta$, defined with respect to the optical axis of the microscope. The intensity distribution is Gaussian ($1/e^2$ diameter $d$ was measured experimentally) according to the method reported in Ref.[43] For oblique incidence, the spot size was Gaussian elliptic, with the long axis $d_x(\theta)$ being angle-dependent and lying in the plane of incidence and the short axis being constant ($d_y = 59\,\mu m$). The peak fluence $F$ was calculated as $F = 8E/\pi \cdot d_x(\theta) \cdot d_y$. It is important noticing that the absorbed fraction of the laser pulse energy is also angle dependent due to the corresponding dependence of the Fresnel reflection coefficient. This was taken into account by formulating an effective fluence $F_{eff} = (1 - R(\theta)) \cdot (8E/\pi \cdot d_x(\theta) \cdot d_y)$. Due to the high reflectivity of Cu at the laser wavelength the effective fluence values quoted in the text are low.

The sample was a commercially available Cu slab with 99.9% purity, mechanically polished afterwards to a final surface roughness given in Table 1 ($N_{eff} = 0$). The sample was mounted on a rotation stage in order to select the irradiation angle $\theta$ and on a motorized three-axis translation stage to position the sample or move it at a user-defined constant speed. Since the laser was operated at a constant pulse repetition rate, the setting of the speed effectively controlled the effective pulse number per unit area incident on the sample. The effective pulse number per unit area for a given spot size and sample speed $v$ was defined as $N_{eff} = d_x(\theta) \cdot 100\,Hz\,/\,v$.

After irradiation, the laser-exposed regions were characterized using a variety of techniques. Optical microscopy was performed with an $NA = 0.9$ and $460\,nm$ illumination, yielding a maximum lateral resolution $R_{xy} < 300\ nm$. The surface topography of the structures was measured with an atomic force microscope operating in tapping mode.

The period of the LIPSS structures has been determined via performing a fast Fourier transform on the optical and SEM micrographs of the fringe structures. The error bar assigned to the obtained period value corresponds to two times the standard deviation of a Gaussian fit to the first order of the FFT.

Acknowledgements:

JSi, JSo, YF and CF acknowledge the Spanish Ministry of Science, Innovation and Universities for financial support through research grant UDiSON (TEC2017-82464-R) and the European Commission for grant LiNaBioFluid (FETOPEN 665337). VG and JASG acknowledge the Spanish Ministerio de Economia y Competitividad for financial support through the grants LENSBEAM (FIS2015-69295-C3-2-P) and NANOTOPO (FIS2017-91413-EXP), and also Consejo Superior de Investigaciones Científicas (INTRAMURALES 201750I039).